\documentclass[twocolumn,showpacs,preprintnumbers,amsmath,amssymb]{revtex4}

\usepackage{graphicx}
\usepackage{dcolumn}
\usepackage{bm}

\newcommand{\ybco}{YBa$_2$Cu$_3$O$_{7-\delta}$ }

\begin{document}


\title{Quasi-static and dynamic order-disorder transition in presence of strong pinning}

\author{A. V. Bondarenko}
 \email{Aleksandr.V.Bondarenko@univer.kharkov.ua}
\author{A. A. Zavgorodniy}
\author{D. A. Lotnik}
\author{M. A. Obolenskii}
\author{R. V. Vovk }
\affiliation{Physical department, V.N. Karazin Kharkov National
University, 4 Svoboda Square, 61077 Kharkov, Ukraine.
}%

\author{Y. Biletskiy}
\affiliation{Department of Electrical and Computer Engineering,
University of New Brunswick, 15 Dineen drive, Fredericton, New
Brunswick, E3B5A3, Canada
}%

\date{\today}

\begin{abstract}
We present the results of an experimental study of vortex dynamics
in non-twinned $YBa_2Cu_3O_{6,87}$ crystal. It is found that
critical currents $J_c$ and $J_{c,dyn}$, which correspond to the
pinning force in the thermal creep and flux flow mode,
respectively, non-monotonically vary with the magnetic field.
However, the minimum in the $J_{c,dyn}(H)$ dependence is observed
in higher fields, compared with the minimum position $H_{OD}$ in
the $J_c(H)$ dependence. Considering that the field $H_{OD}$
corresponds to the static order-disorder transition, this
difference is explained by partial dynamic ordering of the vortex
solid. It is concluded that finite transverse barriers guarantee
finite density of transverse displacements of vortex lines
$u_t\simeq c_La_0$ suitable for preservation of the disordered
state of the moving vortex solid.
\end{abstract}

\maketitle


The interaction of static and dynamic elastic media with chaotic
pinning potential is one the chapters of solid state physics,
which includes dislocations in solids, charge density waves,
Vigner crystals, and vortex lattices (VL's) in Type-II
superconductors. The VL's are the most appropriate objects for the
experimental study of elastic media, because it is easy to change
the strength of pinning potential in superconductors, as well as
the elasticity and motion velocity of VL's. An important feature
of the VL's is the non-monotonous field variation of the pinning
force $F_p$, which is observed in low-T$_c$ (NbSe$_2$
\cite{Bhattacharya93,Higgins96}, V$_3$Si \cite{Gapud03})
middle-$T_c$ (MgB$_2$ \cite{Pissas02, Kim04}), and high-T$_c$
(BiSrCaCuO \cite{Khaikovich96}, YBaCuO \cite{Kupfer98, Pissas00})
superconductors. The increase of the pinning force can be
explained by softening of the elastic moduli of VL's in vicinity
of the upper critical field $H_{c2}(T)$ \cite{Higgins96} or the
melting line $H_m(T)$ \cite{Kwok94} that causes better adaptation
of the vortex lines to the pinning landscape. Some alternative
models \cite{Ertas97, Rosenstein07} suggest formation of an
ordered vortex solid (VS) in low fields, which transforms into a
disordered one in some magnetic field $H_{OD}$, though the nature
of the order-disorder (OD) transition and the mechanism of
increasing the force $F_p$ may be different. These models are
supported by correlation between the field $H_{OD}$ corresponded
to the structural OD transition \cite{Cubbit93} and the onset of
the $F_p$ increase \cite{Khaikovich96} in BiCaSrCuO crystals. An
actual problem of the VS phase is the nature of its ordering under
an increased vortex velocity $v$. The "shaking temperature" model
\cite{Koshelev94} suggests that transverse vortex displacements
$u_t$ induced by the disorder reduce with increased velocity,
$u_t\propto 1/v$; and the increase of the velocity above some
critical value $v_c$ results in a dynamic transition from the
disordered to ordered state. It was later justified
\cite{Giamarchi96} that the increase in $v$ leads to a suppression
of the pinning in the longitudinal (with respect to $\textbf{v}$)
direction only, while pinning barriers remain finite in the
transverse direction. The effect of motion on the transverse
barriers, phase state and pinning force of vortex solid is still
controversial issue, and this subject first of all requires
additional reliable experimental studies. The goal of this work is
experimental study of vortex dynamics in the presence of strong
pinning.

The measurements were performed on detwinned \ybco crystal,
annealed in an oxygen atmosphere at 500$^{\circ}$C for one week.
Such anneal corresponded to an oxygen deficiency $\delta\simeq$
0.13 \cite{Otterlo00} and $T_c\simeq$ 91.8~K. The crystal then was
held at room temperature for 7 days to form clusters of oxygen
vacancies, which reduced the tension of the field $H_{OD}$
\cite{Liang98}. The field variation of the pinning force was
studied through measurement of the current-voltage
characteristics, $E(J)$, using the standard four-probe method with
dc current. The investigated sample had rectangular shape with
smooth surfaces; its dimensions were 3.5$\times$0.4$\times$0.02~mm
with the smallest dimension along the $c$ axis; the current was
applied along the largest dimension; and the distance between the
current and potential contacts, and between the potential contacts
was about 0.5~mm. The measurements were performed at a temperature
of 86.7~K in the field $\textbf{H}\parallel \textbf{c}$.
\begin{figure}
\includegraphics[clip=true,width=3.2in]{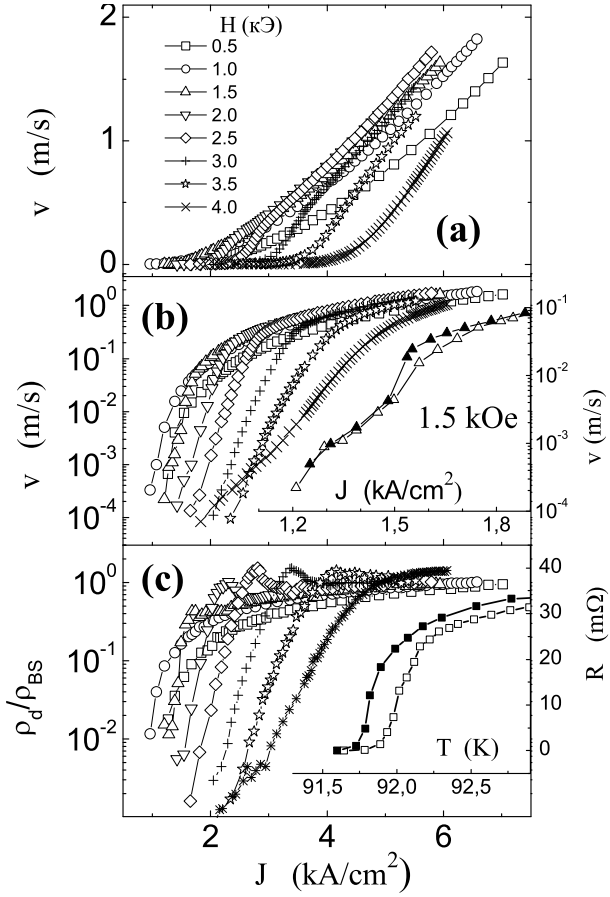}
\caption{\label{fig:1} $E(J)$ curves presented in the linear (a)
and semi-logarithmical scale (b), and $\rho_d(J)$ curves presented
in the semi-logarithmical scale (c). The inset in panel (b) shows
the $E(J)$ dependencies measured upon increase (light symbols) and
decrease (dark symbols) of the current.}
\end{figure}
Fig.~\ref{fig:1} shows the $v(J)=cE(J)/B$ dependencies and current
variation of the normalized dynamic resistance
$\rho_d(J)\equiv[dE(j)/dJ]/\rho_{BS}$, where $\rho_{BS}=\rho_N
B/B_{c2}$\cite{Kupfer98}. At low currents, the electric field
increases exponentially with an increase in current and the
resistance $\rho_d$ is much lower than one. This increase in $v$
and low dynamic resistance indicate the presence of thermally
activated vortex creep. At high currents, the $v(J)$ dependence is
linear and the value of $\rho_d$ is close to 1, indicating the
presence of the flux flow mode. The critical current in the
thermal creep mode $J_E$ can be characterized by the voltage
criteria of $E = 1~\mu$V/cm and $E =100~\mu$V/cm, and the dynamic
critical current $J_{c,dyn}$ can be determined by extrapolating
the linear parts of the $v(J)$ dependence, corresponded to the
flux flow mode, to zero voltage \cite{Kokubo07}. Field variation
of the currents $J_E$ and $J_{c,dyn}$ normalized by their values
in a field of 0.5~kOe are shown in Fig.~\ref{fig:2}a. It is seen
that the currents $J_c$ and $J_{c,dyn}$ start to increase in the
fields above 1.25~kOe and 2.5~kOe, respectively, which are
substantially smaller in comparison with the fields $H_{c2}$ and
$H_m$. Therefore this increase can not be caused by better
adaptation of the vortices to the pinning landscape induced by
softening of the elastic moduli. Obtained field variation of the
currents $J_E$ and $J_{c,dyn}$, and the peculiarities of vortex
dynamics can be explained in frames of the model proposed by Ertas
and Nelson \cite{Ertas97}. It is assumed that the OD transition
occurs when transverse displacements of vortex lines exceed the
value of $c_La_0$, where $a_0\simeq\sqrt{\Phi_0/B}$ is intervortex
distance, $\Phi_0$ is the flux quantum, and $c_L$ is the Lindemann
number. The field is defined by equality of energies
$E_{el}(H_{OD}) = E_p(H_{OD})$, where $E_p$ is the pinning energy,
$E_{el}\simeq c_L^2\varepsilon\varepsilon_0a_0$ is increase of the
elastic energy caused by displacements $u_{t}=c_{L}a_0$,
$\varepsilon$ is the anisotropy parameter, $\varepsilon_0 =
(\Phi_0/4\pi\lambda)^2$ is the line tension of vortex line and
$\lambda$ is the penetration depth. As evident from
Fig.~\ref{fig:2}a, the minimum position does not depend on the
driving force within the creep regime in agreement with
magnetization measurements \cite{Kupfer98,Pissas00}. This means
that the value of ratio $E_{el}/E_p$, and, therefore the energy
$E_p$, is not changed, indicating that minimum in the $J_E(H)$
curve corresponds to static OD transition,
$H_{OD}\simeq$~1.25~kOe.
\begin{figure}
\includegraphics[clip=true,width=3.2in]{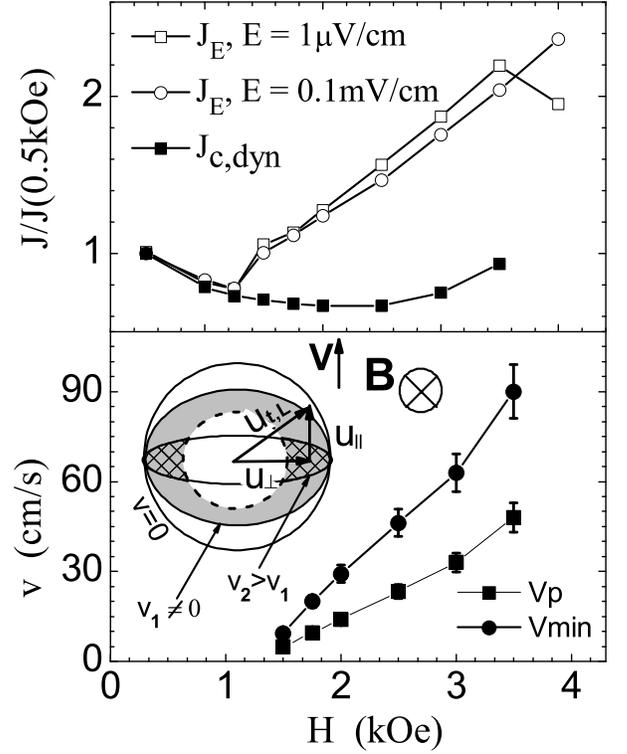}
\caption{\label{fig:2} (a) Field variation of the current
$J_{c,dyn}$ and $J_E$ normalized by their values in a field of
0.5~kOe. (b) Field variation of the velocities $v_p$ and $v_{min}$
correspondent to the peak and minimum position in the $\rho_d(J)$
dependencies, respectively. The inset in panel (b) shows sketch of
the transverse vortex displacement $u_{t,L}$ correspondent to the
Lindemann criteria. Dash and solid circles correspond to the lower
and upper boundaries of the displacements $u_{t,L}$ (see the
text), respectively, in the static VS in magnetic field $H >
H_{OD}$. Dot ellipses show evolution of the maximal displacements
$u_{t,L}$ upon increase of the velocity $v$. Dashed region
corresponds to the displacements $u_{t,L}$ in the dynamic VS. }
\end{figure}
Estimations presented below show that static OD transition in our
sample is caused by vortex interaction with the clusters of oxygen
vacancies rather than with the isolated oxygen vacancies. Indeed,
for the point disorder the pinning energy is
\cite{Blatter94,Ertas97} $E_p\simeq
(\gamma\varepsilon^2\varepsilon_0\xi^4)^{1/3}(L_0/L_c)^{1/5}$,
where $L_0\simeq 2\varepsilon a_0$ is the length of longitudinal
fluctuations, $L_c\simeq\varepsilon\xi (J_0/J_d)^{1/2}$ is the
correlation length, $J_0=4c\varepsilon_0/3\sqrt{3}\xi\Phi_0$ is
the depairing current, $\xi$ is the coherence length, and
$\gamma\simeq(J_c\Phi_0/c)^2L_c$ is the disorder parameter. Using
realistic for the \ybco superconductor parameters ($\lambda$ = 500
nm, $\xi$ = 4 nm, and $\varepsilon$ = 1/7) and experimental value
of the depinning current $J_{c,dyn}<$ 5 kA/cm$^2$ we obtain the
energy $E_p <$ 2$\cdot$10$^{-16}$ erg, which is about 25 times
smaller compared to the elastic energy $E_{el}\simeq
c_L^2\varepsilon\varepsilon_0a_0\simeq$ 5$\cdot$10$^{-15}$ erg
estimated for the $c_L$ = 0.2 and $H_{OD}$ = 1.25 kOe. The pinning
energy induced by vortex interaction with the clusters of oxygen
vacancies equals the condensation energy
$U_c\approx(H_c^2/8\pi)V_{cl}$, where
$H_c=\Phi_0/2\sqrt{2}\pi\lambda\xi$ is the thermodynamic critical
field and $V_{cl}$ is the volume of clusters. For spherical
clusters with radius $r\simeq\xi$ we obtain the energy $E_p\simeq
U_c\approx$ 10$^{-14}$ erg, which is suitable for occurrence of
the OD transition in the field of 1.25 kOe.

As it is shown in Fig.~\ref{fig:2}b, minimum in the $J_{c,dyn}(H)$
curve occurs in a field of 2.5~kOe, which is about two times
exceeds the value of $H_{OD}$. Also, above the minimum position,
the current $J_{c,dyn}(H)$ increases with the field more gradually
in comparison with increase of the current $J_E$ above the OD
transition. This difference can be explained by suppression of the
longitudinal and conservation of the transverse pinning barriers,
as it was theoretically predicted in \cite{Giamarchi96}. In frames
of the "shaking temperature" model \cite{Koshelev94}, this means
conservation of the transverse (with respect to vector
$\textbf{v}$) $u_{\perp}$ and reduction of the parallel
$u_{\parallel}\propto 1/v$ component of the displacements $u_t =
\sqrt{(u_{\parallel})^2+(u_{\perp})^2}$ with increased velocity
$v$. In magnetic field $\textbf{H}\parallel \textbf{c}$ and in
presence of the chaotic pinning potential, spatial distribution of
the displacements is isotropic; and in the field $H > H_{OD}$, the
displacements $u_{t,L}$, which correspond to the Lindemann
criteria, fall in the interval $c_La_0(H_{OD}) > u_{t,L} >
c_La_0(H)$, as it is shown schematically in the inset of Fig. 2b.
Density of the displacements (the number of vortex displacements
$u_{t,L}$ per unit length of vortex line) $n_{t,L}$ is
proportional to the area of ring confined by the upper (solid
circle) and lower (dashed circle) boundary of the displacements
$u_{t,L}$. Reduction of the component $u_{\parallel}$ with
increased velocity $v$ leads to reduction of the upper boundary
(dotted lines for velocities $v_2 > v_1\neq 0$), and thus to
reduction of the density $n_{t,L}$. It is important, that for any
finite velocity $v$ the component $u_{\parallel}$ is finite, and
thus the cross-hatched area at the diagram, which corresponds to
the displacements $u_{t,L}$, and the density $n_{t,L}$ is also
finite. Increase of the field reduces the lower boundary of the
displacements $u_{t,L}$, and therefore the density $n_{t,L}$
increases.

Considering that the displacements $u_{t,L}$ produce the
dislocations in the VS phase, and increase of the density
$n_{t,L}$ results in an increase of the current $J_{c,dyn}$
\cite{Bondarenko08a}, the field variation of the currents $J_E$
and $J_{c,dyn}$ can be explained in the following way. In low
fields, the ordered VS phase, which is characterized by the
absence of dislocation and realization of the 1D pinning, is
formed, and the currents $J_E$ and $J_{c,dyn}$ decrease with
increased field due to enhancement of the vortex-vortex
interaction, making difficult to fit the vortices in the pinning
landscape. Above the OD transition, the VS phase contains
dislocations that results in occurance of the 3D pinning
\cite{Ertas97}, and thus the current $J_E$ increases at the
transition point $H_{OD}$ due to dimensional crossover in the
pinning \cite{Kes86, Brandt86}. Further increase of the current
$J_E$ with magnetic field is caused by increase of the density
$n_{t,L}$, as it was found in \cite{Bondarenko08a}. The density
$n_{t,L}$ is smaller in the moving VS phase than in the static VS
phase, but it is finite and increases with the field. Therefore,
the $J_{c,dyn}(H)$ dependence is determined by competition between
decrease of the pinning force caused by enhancement of the
vortex-vortex interaction and increase of the pinning force
associated with increase of the density $n_{t,L}$. In our
measurements, the former mechanism dominates in magnetic fields
$H\leq$~2~kOe, while the last one dominates in the fields
$H\geq$~3~kOe.

Proposed interpretation agrees with numerical simulations of 2D
\cite{Faleski96, Moon96, Olson00, Kolton99} and 3D
\cite{Otterlo00} VL's in the presence of strong pinning. First, it
was shown that in the flux flow mode the disordered state of the
VL's is preserved \cite{Faleski96, Kolton99, Otterlo00, Olson00},
and the transverse barriers remain finite \cite{Moon96, Olson00}.
Second, the $v(J)$ curves cross one another near the OD transition
\cite{Otterlo00}. Third, our interpretation implies that
cross-hatched area in the diagram collapses to a segment at
$v\rightarrow\infty$, indicating that moving VS can be ordered in
agreement with conclusion in \cite{Otterlo00}. Finally, the onset
of ordering of the moving VS phase is manifested as a peak in the
$\rho_d(J)$ curves, and the end of ordering corresponds to value
of the resistance $\rho_d(J)$=1 \cite{Faleski96, Kolton99}, and in
our measurements peak in the $\rho_d(J)$ curves appears in the
fields $H > H_{OD}$. Following computer simulations, we determined
the field variation of the velocities $v_p$ and $v_{min}$, which
correspond to the peak and minimum positions in the $\rho_d(J)$
curves respectively. As it is shown in Fig. 3b, the velocity $v_p$
and the difference $\Delta v = v_{min} - v_p$ increase with the
field indicating that the critical velocity of the ordering as
well as the interval of velocities $\Delta v$, in which the
ordering realizes, increase with the field. This behavior is
plausible considering that the lower boundary of the displacements
$u_{t,L}$ decreases with the increased field that requires higher
$v$'s to decrease the amplitude below this boundary. Also, the
difference between the upper and lower boundary of the
displacements $u_{t,L}$, $\Delta u = c_L[a_0(H_{OD})-a_0(H)]$,
increases with the field that results in increase of the
difference $\Delta v$.

Our interpretation allows explaining occurrence of the hysteresis
effect in the curve $v(J)$ measured with the increased and
decreased current in a field of 1.5~kOe, and absence of the
hysteresis effect below and quite above the OD transition. Indeed,
in close vicinity to the OD transition, $(H/H_{OD} - 1) << 1$, the
density $n_{t,L}$ in the dynamic VS is much smaller than in static
VS, and small increase of the velocity $v$ leads to dynamic
transition into the ordered state. In this case the "shaking
temperature" model predicts the hysteresis effect, which reflects
the "overheated state" of the ordered dynamic VS. The decrease in
density $n_{t,L}$ quite above the OD transition is not dramatic,
and transition from strongly disordered static VS to less
disordered dynamic VS occurs in a wide interval of velocities
$\Delta v$ without hysteresis. It is important to notice that the
$E(J)$ curves measured after zero field cooling coincide with the
$E(J)$ curves measured after non zero field cooling, that
indicates the absence of metastable states in the VS. This agrees
with experimental studies of the YBaCuO crystals: the metastable
states exist in vicinity of the vortex sold - vortex liquid
transition, but they disappear below this transition
\cite{Fendrich96}.

Recent quantitative theory of the dynamic VS by Rosenstein and
Zhuravlev \cite{Rosenstein07} predicts jump-like increase of the
pinning force at the OD transition. It is evident that this theory
does not describe our results because increase of the currents
$J_E$ and $J_{c,dyn}(H)$ occurs in different fields, and field
variation of the currents does not show the jump-like increase.

The obtained field variation of the currents $J_E$ and
$J_{c,dyn}$, occurrence of the hysteresis effect in close vicinity
to the OD transition, and absence of the metastable states in the
VS are different from that in superconductors with weak bulk
pinning. For example, in crystals NbSe$_2$ \cite{Henderson96} and
MgB$_2$ \cite{Kim04}, the current $J_{c,dyn}$ increases in a
jump-like manner at the OD transition
\cite{Henderson96,Henderson98}, and the hysteresis effect occurs
in rather wide interval of magnetic fields and it is caused by
presence of the metastable states in the VS
\cite{Henderson96,Kim04}, which are induced the effect of surface
barriers \cite{Paltiel00}. The surface barriers in the NbSe$_2$
\cite{Banerjee00} and MgB$_2$ \cite{Pissas02} cause asymmetry of
the magnetization loops, and this asymmetry reflects a difference
in the barriers for vortex entrance and exit of samples
\cite{Burlachkov93}. The magnetization loops of the YBaCuO
crystals are symmetric indicating negligible effect of the surface
barriers. Therefore obtained field variation of the current $J_E$
corresponds to equilibrium quasistatic VS.

In conclusion, we determined field variation of the critical
currents in the quasistatic and dynamic vortex solid. The currents
non-monotonously vary with the field, but minimum position in the
$J_{c,dyn}(H)$ dependence is shifted to higher fields in
comparison with the minimum in the $J_E(H)$ dependence. The
difference is interpreted by partial ordering of the vortex solid
with increased vortex velocity. The disordered state of the
dynamic vortex solid is attributed to preservation of finite
transverse pinning barriers that guarantees presence of the
transverse vortex displacements suitable for formation of
dislocations. This interpretation allows explaining observed
increase of the critical current of the dynamic ordering.

\bibliography{/bondarenko/tex.sample/paper}

\end{document}